\documentclass[fleqn,12pt]{article}

\newcommand{\AmS}{{\protect\the\textfont2
  A\kern-.1667em\lower.5ex\hbox{M}\kern-.125emS}}

\title{Present and Future Experiments
in Non-equilibrium Reactor Antineutrino Energy Spectrum}

\author{V. I. Kopeikin, L. A. Mikaelyan \\
\\
Russian Research Center "Kurchatov Institute"}
 
\begin{document}

\date{}
\maketitle


\begin{abstract}
Considerable efforts that have been undertaken in the recent years in low energy antineutrino experiments require further 
systematic investigations in line of reactor antineutrino spectroscopy as a metrological basis of these experiments. We consider 
some effects associated with the non-equilibrium of reactor $\bar{\nu_e}$-radiation and residual $\bar{\nu_e}$-emission 
from spent reactor fuel in contemporary $\bar{\nu_e}$ experiments.
\end{abstract}

\section*{Introduction}

In the past decade revolutionary progress in the low-energy antineutrino detection technique has been made. Unprecedented 
improvements in sensitivity, precision and low-background detection of small energy deposition at reactor experiments required 
exact knowledge of the non-equilibrium reactor $\bar{\nu_e}$-spectrum, see e.g. overview [1]. Recent registration of the 
geoneutrinos with KamLAND [2] in the relatively big reactor antineutrino flux calls for further accurate studies of the reactor 
$\bar{{\nu}_e}$-spectrum.

We consider some effects associated with the non-equilibrium behavior of reactor $\bar{\nu_e}$-radiation during operating 
and shutdown periods and $\bar{\nu_e}$-emission from spent nuclear fuel stored near reactor on the following 
$\bar{\nu_e}$ experiments and projects:

\begin{description}
\item [\rm(i)] Searches for neutrino magnetic moment.
\item [\rm(ii)] Sensitive searches of the mixing angle $\theta_{13}$ in oscillation experiments.
\item [\rm(iii)] Observation of U- and Th antineutrinos from the Earth.
\end{description}

\section{Non-equilibrium reactor antineutrino\\ spectrum}

1.1. We consider here the widespread pressurized light water power reactors (PWR), which operate for 11 months, followed 
by shutdown of 1 month for replacing 1/3 of the spent nuclear fuel, which is kept for some years in water spent fuel pool (SFP) 
near the reactor. The average relative fuel composition of a reactor core is (in fissions for fissile isotopes):

$$
\bar\alpha_5=0.58\ {\rm for\ ^{235}U}, \quad \bar\alpha_9=0.30\ {\rm for\ ^{239}Pu},\qquad \qquad \qquad \qquad \qquad 
\;
$$

\begin{equation}
\bar\alpha_8=0.07\ {\rm for\ ^{238}U}, \quad \bar\alpha_1=0.05\ {\rm for\ ^{241}Pu}.
\end{equation}

Reactor $\bar{\nu_e}$-spectrum is broadly distributed over energies up to about 10 MeV, with peaks at $\sim$ 0.3 MeV. 
From the start of reactor, the $\bar{\nu_e}$-spectrum begins to evolve slowly towards equilibrium; after reactor is shut 
down, the $\bar{\nu_e}$-spectrum falls down for a long time. There are four sources of reactor $\bar{\nu_e}$-spectrum 
evolution: (a) accumulation and decay of fission products of each of the four fuel isotopes (1); (b) changes of the reactor fuel 
composition caused by burn up effects; (c) beta-decay of the $\rm^{239}U\to^{239}Np\to^{239}Pu$ chain produced via 
neutron radiation capture in $\rm^{238}U$; and (d) neutron captures by fission products. The calculation of the non-equilibrium
 reactor $\bar{\nu_e}$-spectrum is presented in [3].
 
1.2. The most precision information on reactor $\bar{{\nu}_e}$-spectrum was obtained for energies above the inverse 
beta-decay reaction threshold $E_{\rm th} \approx 1.8 {\rm\; MeV}$ [4,5]:

\begin{equation}
\qquad \qquad \qquad \qquad \bar{\nu_e} + p \rightarrow n + e^{+}
\end{equation}

Conventionally used procedure of finding the reactor zero approximation $\bar{\nu_e}$-spectrum $\rho_0(E,t)$, 
in units of $\bar{\nu_e} \cdot \rm MeV^{-1} \cdot fission^{-1}$, is known:

\begin{equation}
\qquad \qquad \qquad \rho_0(E,t) = \sum \alpha_i(t) \cdot \rho^i_0(E)
\end{equation}
where $\alpha_i(t)$, $\Sigma \alpha_i(t)$ = 1, are the time-dependent contributions and $\rho^i_0(E)$ are the 
time-independent energy $\bar{\nu_e}$-spectra ($E > 1.8$ MeV) of the fissile isotopes (1). For all but $^{238}\rm U$ is used 
the ILL collaboration "converted" $\bar{\nu_e}$-spectra, which correspond to only $\sim$ 1-day fuel irradiation time [4]. 
For $^{238}\rm U$ whose contribution is low calculation method is used.

According to zero approximation, the $\bar{\nu_e}$-emission for region $E > 1.8$ MeV falls down to zero within 1-day after 
reactor shutdown.

We calculate corrections to the $\rho_0(E,t)$, which associate with (a) long-lived fission products accumulation and (b) 
neutron captures by fission products in reactor core, see Fig. 1a. Among long-lived fission products e.g. are:

$$
 ^{144}{\rm Ce} \ ( T_{1/2} = 285 \ {\rm d}) \to  ^{144}\!\!{\rm Pr} \ ( T_{1/2} = 17 \ {\rm m}, E_{\rm max} = 2.997 \ 
{\rm MeV}) \quad
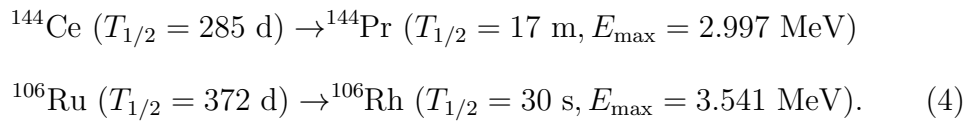
$$

\begin{equation}
^{106}{\rm Ru} \ ( T_{1/2} = 372 \ {\rm d}) \to  ^{106}\!\!{\rm Rh} \ ( T_{1/2} = 30 \ {\rm s}, E_{\rm max} = 3.541 \ 
{\rm MeV}). \qquad
\end{equation}
Also we calculate the residual $\bar{\nu_e}$-radiation from stopped reactor and SFP, see Fig. 1b. These corrections to the 
$\rho_0(E,t)$ and residual $\bar{\nu_e}$-radiation all together form an ignored non-equilibrium effect in reactor 
$\bar{\nu_e}$-spectrum and can play a significant role in the contemporary neutrino experiments.

1.3. The spectrum of reactor antineutrinos below 1.8 MeV has been studied in a systematic way from the middle of 1990-th 
[6,7,3]. The first data of this part of the $\bar{\nu_e}$-spectrum are presented in [7]. Now we show in Table 1 our latest data 
of the $\bar{\nu_e}$-spectrum $\rho(E)$ at the middle of reactor operating period $t \approx$ 0.5 year and for $E <$ 3.5 
MeV. We calculated the $\bar{\nu_e}$-spectrum below 2 MeV; our results between 2 MeV and 3.5 MeV are based on zero 
approximation procedure (3) with calculated corrections (see Section 1.2 and Fig. 1a). The spectrum above 3.5 MeV can be 
approximated by (3).

\begin{table}[htb]
\caption{Reactor antineutrino spectrum $\rho (E)$ in the middle of operating period (see text) for energies $E \le $ 3.5 MeV 
in units of ($\bar{\nu_e} \cdot \rm MeV^{-1} \cdot fission^{-1}$)}
\vspace{3pt}
\begin{tabular}{c|c|c||c|c|c}
\hline
$E$, MeV & $\rho(E)$ & $\delta\rho^*,\%$ & $E$, MeV & $\rho(E)$ & $\delta\rho,\%$\\
\hline
0.010 & .6548(-1)$^{**}$&7&0.500& .2997(+1) & $\div$\\
0.020 & .2545(+0) & $\div$ & 0.600 & .3008(+1) & $\div$\\
0.035 & .7167(+0) & $\div$ & 0.800 & .3206(+1) & $\div$\\
0.040 & .3813(+0) & $\div$ & 0.900 & .3235(+1) & $\div$\\
0.070 & .1054(+1) & $\div$ & 1.000 & .3095(+1) & $\div$\\ 
0.100 & .1776(+1) & $\div$ & 1.185 & .2777(+1) & $\div$\\
0.140 & .3051(+1) & $\div$ & 1.190 & .2204(+1) & 7        \\
0.160 & .3785(+1) & $\div$ & 1.250 & .2003(+1) & 5        \\
0.165 & .3161(+1) & $\div$ & 1.300 & .1777(+1) & $\div$\\
0.180 & .3627(+1) & $\div$ & 1.500 & .1595(+1) & $\div$\\
0.215 & .4503(+1) & $\div$ & 1.700 & .1485(+1) & $\div$\\
0.230 & .3911(+1) & $\div$ & 1.800 & .1421(+1) & $\div$\\
0.280 & .4803(+1) & $\div$ & 1.900 & .1350(+1) & 5       \\
0.330 & .5728(+1) & $\div$ & 2.000 & .1270(+1) & 3       \\
0.335 & .4363(+1) & $\div$ & 2.250 & .1078(+1) & $\div$\\
0.350 & .4159(+1) & $\div$ & 2.500 & .8808(+0) & $\div$\\
0.390 & .4510(+1) & $\div$ & 2.750 & .7429(+0) & $\div$\\
0.400 & .4264(+1) & $\div$ & 3.000 & .6115(+0) & $\div$\\
0.435 & .4618(+1) & $\div$ & 3.250 & .5083(+0) & $\div$\\
0.440 & .2810(+1) & $\div$ & 3.500 & .4119(+0) & 3        \\
\hline
\end{tabular} \\
\vspace{5pt}
$^*$ Evaluation of the error corridor (68\% CL)\\
$^{**}$ .6548(-1) = $0.6548 \cdot 10^{-1}$
\end{table}

\section{Searches for the neutrino magnetic moment}

It should be recalled that the differential cross section for magnetic $\bar{\nu_e}e$- scattering $d\sigma^{\rm M} / dT$ behaves 
as $\sim 1/T$, ($T$ is recoil-electron kinetic energy), whereas the cross section for weak $\bar{\nu_e}e$- scattering 
$d\sigma^{\rm W} / dT$ tends to a finite value when $T \to 0$. In order to observe the neutrino magnetic moment 
$\mu_\nu$ at a level of $10^{-11}\mu_ B$ measurements ought to be performed in the region of $T < 10$ keV [1,6].

A constraint $\mu_\nu < 1\times 10^{-10}\mu_B$ (68\% CL) was derived from TEXONO collaboration measurements 
with HPGe detector $\sim$ 1 kg in the reactor flux of $6 \times 10^{12}\ \bar{\nu_e}\ {\rm cm^{-2}s^{-1}}$ [8]. Further 
measurements are pursued. The GEMMA experiment [9] with HPGe crystal of 2 kg and NESSI experiment [10] using 
a 80 kg semiconductor silicon detector aimed to reach a sensitivity of $\mu_\nu \sim 3 \times 10^{-11}\mu_B$.

The relevant quantities for $\bar{\nu_e}e$- scattering experiment are the folded weak (W) and magnetic (M) integral cross 
sections $I^{\rm W, M}$ in the interval ($1-T$)\,keV:

\begin{equation}
I^{\rm W, M} = \int_{\rm 1 keV}^T dT\int_{E_{\rm min}}^\infty dE \ \frac{d\sigma^{\rm W,M}(T,E)}{dT}\ \rho(E,t).
\end{equation}

We calculated the time variations of the W- and M ($\mu_\nu = 3 \times 10^{-11}\mu_B$) $\bar{\nu_e}e$- scattering 
effects both during reactor operating (Fig. 2a) and shutdown (Fig.\,2b) periods. The W- and M effects increase during 
operating period $\sim 1\rm\ day \div\,1\ year$ reach about 20\%, typical residual W- and M effects during shutdown period are 
from 5\% to 20\%. They are associated with the non-equilibrium of reactor $\bar{\nu_e}$-spectrum and should be 
taken into account.

\section {Sensitive measurement of the mixing\\ angle $\vartheta_{13}$}

CHOOZ collaboration used one spectrometer stationed at $\sim$ 1 km from reactor(s) and measured positron spectrum and 
rate of reaction (2) and obtained the upper limit of $\vartheta_{13}$ [11]: 

\begin{equation}
\sin^2{2\vartheta}_{13} \le 0.14 \;(90\% \;{\rm CL \;for} \;\Delta m^2 = 2.5 \times 10^{-3}\;{\rm eV^2}).
\end{equation}

For radical increasing of this constraint an idea of one reactor $-$ two identical detectors (near and far) was proposed [12] 
and elaborated [13]. The near detector measures $\bar{\nu_e}$-spectrum, while the far detector looks for a deformation of 
this spectrum due to oscillations. Searches for the oscillation parameters are based on an analysis of small deviations of the ratio 
of positron spectra in the far and near detectors from the constant value. The results of this purely relative method are 
independent of the exact knowledge of reactor power, $\bar{\nu_e}$-spectrum, burn up effects $\ldots$ However even for this 
simple layout, see Fig. 3, a distortion $\sim 1.5\%$ can be induced by $\bar{\nu_e}$-radiation from SFP, see Fig. 4.

Now several projects (D-CHOOZ, Braidwood, KASKA $et \ al.$) are considered with some number reactors (from 2 to 7) 
and detectors. For these complicated layouts it must be taken into account both SFP and residual reactor $\bar{\nu_e}$-
radiations and corrections to the $\rho_0(E,t)$, (see Fig. 1a,b).

\section {Observation of geoneutrinos}

Uranium and Thorium geoneutrinos are detected in a large liquid scintillation ($\rm CH_2-$) spectrometer via reaction (2). 
The annihilation quanta are absorbed and positron energy release $E_{vis}$ is related with positron kinetic energy $T$ as:

\begin{equation}
\qquad \qquad \qquad \qquad E_{vis} \approx T+1.02\; \rm(MeV)
\end{equation}

The sensitivity to geoneutrinos is limited by $\bar{\nu_e}$ background from nuclear reactors in the vicinity of the detector. 
The goal of this Section is to consider the role of ignored non-equilibrium effect in reactor $\bar{\nu_e}$-spectrum 
(see Section\,1.2) as a background in searches for geoneutrinos.

We concentrate here on two locations: the Kamioka Laboratory (2700 m.w.e.), where the KamLAND detector is in operation, 
and Baksan Observatory (4700 m.w.e.) as a possible site for developments in geoneutrino physics [14]. We show expected 
positron spectra from geoneutrinos and from reactor- and SFP antineutrinos in spectrometer of  $10^{32}$ protons target 
($\sim$\,1000 ton) with 1 year exposition and efficiency $\varepsilon = 100\%$, see Fig. 5. 
The reactor effect is divided here into two parts: (a) from zero approximation $\bar{\nu_e}$-emission (3), and (b) from ignored 
non-equilibrium $\bar{\nu_e}$-radiation. As can you see in Fig. 5 ignored reactor effect is situated in the energy range of 
geoneutrinos\linebreak ($E_{vis} <$ 2.5 MeV).

At Baksan Observatory the ratio of reactor effect for $E_{vis} <$ 2.5 MeV to U+Th geoneutrinos expected effect is 
$\sim 1/5$, whereas at Kamioka Laboratory is $\sim 5$. Ignored reactor effect at Baksan Observatory is negligible, whereas at 
Kamioka (6.4 events, our calculation) is $\sim 18\%$ of the expected U+Th geoneutrinos effect (36 events according to [15]) 
and approximately equal effect from Th geoneutrinos (7.5 events [15]).

\section*{Conclusion}

We have calculated residual $\bar{\nu_e}$-emission from stopped reactor and spent fuel pool and also found corrections to 
conventionally used reactor $\bar{\nu_e}$-spectrum. It has been shown that these usually ignored features in reactor 
$\bar{\nu_e}$-emission can play a significant role in planning and analyzing neutrino experiments.

\section*{Acknowlegments}

This work was supported by the Russian Foundation for Basic Research (project no 06-02-16024) and by Leading Scientific 
School grant.

\end{document}